\documentclass[prb,twocolumn,showpacs]{revtex4}
\usepackage{graphicx}
\begin{document}
\title{SOS model of overlayer induced faceting}
\author{ Czes{\l}aw Oleksy}
\email{  oleksy@ift.uni.wroc.pl} 
\affiliation{Institute of 
Theoretical Physics, University of Wroc{\l}aw,\\
 Plac Maksa Borna 9, 50-204 Wroc{\l}aw, Poland }
\date{March 4, 2003}

\begin{abstract}
A  solid-on-solid model is proposed to describe faceting of 
bcc(111) metal surface induced by a metal overlayer. It is shown 
that the first order  phase transition occurs between faceted  
\{211\} or \{110\}  and disordered phases. The ordered phases 
consist of large 3-sided pyramids with \{211\} facets or \{110\} 
facets. It is shown that the high-temperature disordered phase 
has not planar bcc(111) structure but faceted disordered 
structure. Hysteresis effects were observed when the system was 
warmed above the transition temperature and then cooled down.  
Temperature dependence of LEED patterns for faceted and 
disordered phase  are calculated in kinematic approximation.

\end{abstract}

\pacs{68.35.Rh, 68.43.De, 64.60.Cn}
 \maketitle

\section{Introduction}

Recent experiments   \cite{mad99a, mad99b, mad96, mad94, mad97, 
mad95, song95} have demonstrated that surfaces such as W(111) and 
Mo(111) covered by a  single physical monolayer of certain metal, 
e.g., Pd, Pt, undergo massive reconstruction from planar 
morphology  to microscopically  faceted surface after annealing 
to $T>700 K$. The reconstructed  surface consists of 3-sided 
pyramids with mainly $\{211\}$ facets, and pyramid dimensions 
range from $\sim 1$ to 100 nm. It has been shown that facets are 
composed of substrate atoms, and the monolayer of adsorbate 
remains on the outermost surface layer during the faceting 
transformation. Another type of massive reconstruction  has been 
very recently found  in STM and LEED experiment for Pd  on 
Ta(111) system \cite{szuk02}. The reconstructed surface consists 
of    $\{011\}$  facets which form  large triangular pyramids. 
The third type of reconstructed structure, coexistence of small 
$\{011\}$ facets with large $\{211\}$ facets,  has been observed 
\cite{mad99b}  in Pd on W(111) for prolonged annealing time (for 
short annealing time only $\{211\}$ pyramids occurred). An 
important step in understanding the thermal stability of 
reconstructed surfaces was LEED experiment performed in high 
temperatures for Pd on Mo(111) by Song et al.\ \cite{song95}. 
They demonstrated existence of     reversible phase transition  
between faceted $\{211\}$ and planar phases and they found that 
this  transition has a large hysteresis, i.e., the transition 
temperature in cooling cycle ($T\approx 830$) is lower than that 
in the heating cycle ($T\approx 870$).

In theoretical studies of overlayer-induced faceting 
\cite{mad99b,leung97,leung98}  the first principle method have 
been used to calculate the surface formation energy of fcc metals 
(Pd, Pt, Au, Ag, Cu) adsorbed on  Mo and W. Results of these 
calculation performed for  pseudomorphic adsorbate overlayer on 
(111), (211), and (011) flat surfaces  show that (111) surface 
becomes unstable at the coverage of one physical monolayer, where 
the energy of (211) orientation has the lowest value. However, 
such energy calculation (at T=0) are not sufficient to explain 
why some of these metals do not induce faceting, e.g., 
Ag/Mo(111). On the other hand, the first principle calculation 
confirmed that critical coverage to induce  faceting  is 
approximately  equal to one physical monolayer (PML) and that the 
growth mode is Stransky-Krastanov as the surface energy increases 
at coverage  higher than 1 PML. 

Theoretical studies of surface reconstruction and surface phase 
transitions  in bimetalic system are complicated problems mainly 
due to occurrence of long range many-body interactions. One of 
approaches to such problems is application of simple 
solid-on-solid models 
\cite{beijeren77,tosat95,beijeren95,kaseno97},   in which the 
crystal is represented by two-dimensional array of columns. They 
were employed to study roughening transition 
\cite{beijeren77,tosat95}, missing-row reconstruction 
\cite{tosat96},  growth of the surface \cite{tosat95},  surface 
diffusion \cite{tosat95,sieradz95}, adsorption \cite{kaseno97}, 
phase transitions in two component crystal  \cite{beijeren95}, 
etc.

In this paper we introduce a simple SOS model for bcc(111) 
surface  covered by one physical monolayer of adsorbed atoms to 
describe surface reconstruction and phase transitions. Using 
Monte Carlo simulation we study change of the surface structure 
during the heating and  cooling  processes,   phase transitions 
to a faceted phases, phase diagram,  LEED patterns and 
temperature dependence of diffracted intensity    calculated in 
kinematic approximation.

\section{The SOS model}\label{s2}

To study faceting in bimetalic system  one needs to know 
interaction potentials. There are many-body potential for Mo-Mo 
and W-W interactions derived by Finnis and Sinclair \cite{finnis} 
but to our knowledge there is only Ni-Mo many-body potential 
\cite{zhang98}  for interaction between (Mo, W) and fcc metals. 
However in Ni/Mo(111) system the faceting does not occur 
\cite{mad96}. Therefore we are going to study overlayer induced 
faceting using a simple solid-on-solid (SOS) model which describe 
a surface formation energy in agreement with result of the first 
principle  calculation.

 In order to construct a  model of overlayer-induced  faceting  
on bcc(111) surface (BCCSOS) we are taking into account some 
experimental evidences. There is a critical coverage, 
approximately equal to 1PML, to induce  faceting on   W(111) and 
Mo(111) \cite{mad94}. When the   coverage  exceeds 1PML, extra  
adsorbate atoms form 3D clusters on the top of faceted surfaces. 
Hence to simplify problem it will be  assumed that coverage is 
constant and  equal to 1PML. Such coverage is equivalent to   3,  
2, and 1 geometric monolayers of adsorbate on bcc  (111), (211), 
and (011) surfaces, respectively. A second assumption is that   
substrate atoms (B)  can take position in discrete bcc lattice. 
It has been shown \cite{mad99b},  in the case of Pd/W(111), that 
both $\{011\}$ facets and  $\{112\}$ facets  of W are covered by  
pseudomorphic monolayer of Pd. Therefore we assume that  
adsorbate atoms (A) are also located in positions of the bcc 
structure. There are three types of interaction energies in this 
SOS model: adsorbate-adsorbate interaction $\epsilon^{AA}_{i}$, 
adsorbate-substrate interaction $ \epsilon^{AB}_{i}$,   and 
substrate-substrate interaction  $\epsilon^{BB}_{i}$, with the  
range of interaction up to forth neighbors  (i =1, 2, 3, 4). 
\begin{figure}
\centering
\includegraphics[width=6cm]{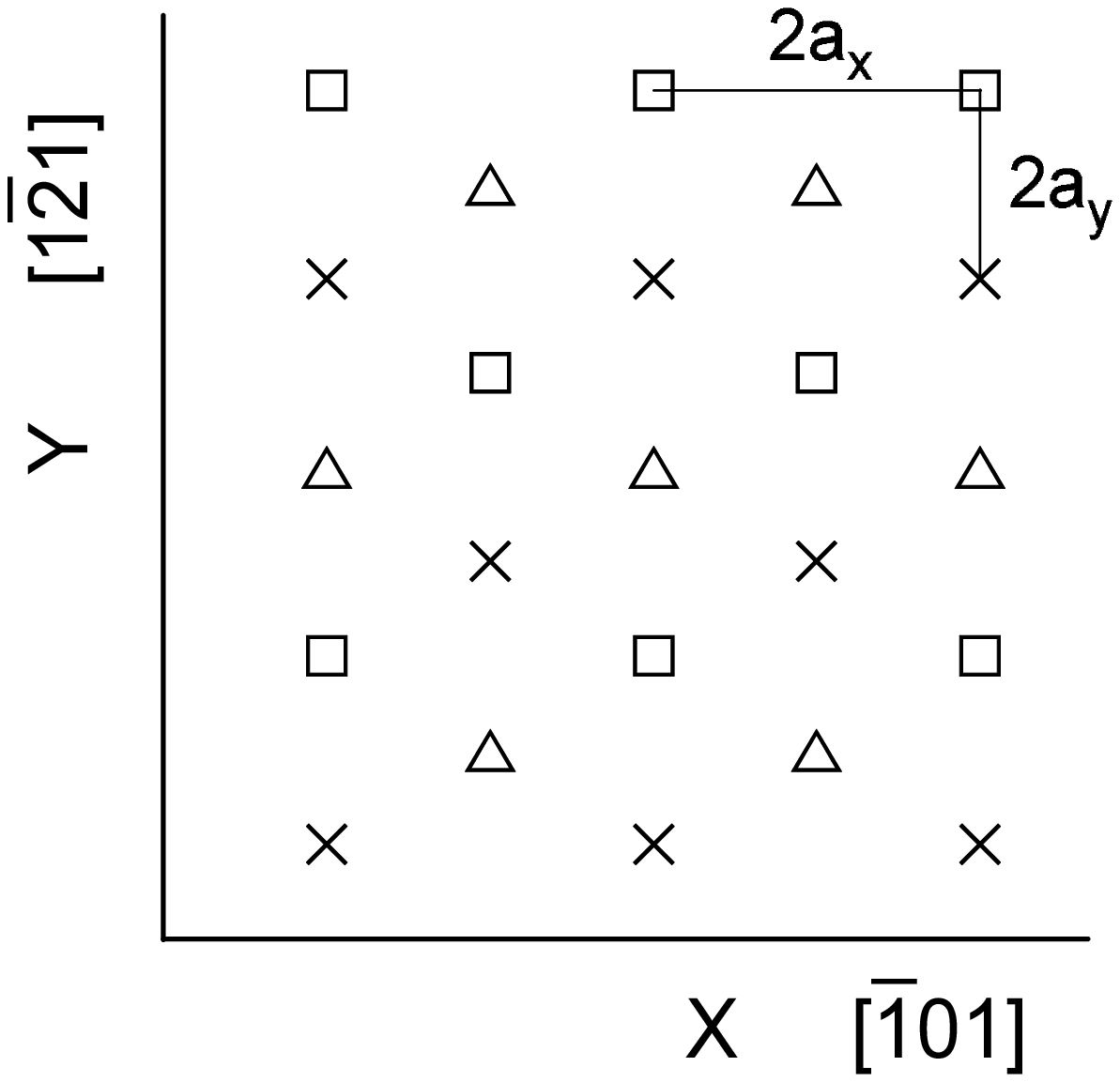}
\caption{ Schematic view of the bcc(111) surface. Sites of the 
sublattices $l=0, 1, 2$ are denoted by squares, triangles, 
crosses, respectively. The Z axis is normal to the plane and 
$a_x=a\sqrt{2}/2$, $a_y=a\sqrt{6}/6$.} \label{figure1}
\end{figure} 

It is convenient to chose  the coordinate system with  the $Z$ 
axis  parallel to $[1 1 1]$ direction and   $X$, $Y$ axes   
along  $[\bar{1} 0 1]$,  $[1 \bar{2} 1]$ directions in the (111) 
plane. Atoms along the closed packed direction ( parallel to the 
z axis) form  columns which positions  in the (x, y) plane  are 
described by three triangular sublattices $l=0, 1, 2$ shown in 
Fig.\ \ref{figure1}. Column height $z_i$ at site {\em i} in the 
{\em l}th sublattice is defined as $z_i = h_i a\sqrt{3}/6$ where  
{\em a} is bcc lattice constant,  $h_i=3 n_i +l$ and  $n_i$ is 
the number of atoms in  {\em i}th column. The assumption on 
constant coverage equal to 1PML means that there is   exactly one 
A atom in each column  placed on its top. As we are going to 
study reconstruction only, not desorption  or crystal melting,  
the typical restriction of  SOS models on columns heights will be 
assumed. The $h_i$ difference  between nearest neighbor sites are 
forced to be $\pm 1, \pm 2$. 

Let us define the Hamiltonian as surface formation energy of A 
material on B (111) surface
\begin{equation}\label{esurf}
 H  =E_{A/B}(N_A, N_B) -E^{bulk}_{A}(N_A)-E^{bulk}_{B}(N_B),
\end{equation}
where $E_{A/B}(N_A, N_B)$  denotes energy of $N_A$  atoms A on 
$bcc(111)$ surface consisting  of $N_B$  atoms B,  
$E^{bulk}_{A}(N_A)$, $E^{bulk}_{B}(N_B)$  are energies of A, B 
atoms in their bulk environments.   Now taking into account above 
assumptions we can express Hamiltonian Eq.\ (\ref{esurf}) in 
terms of column heights $h_i$  in the following form

\begin{equation}\label{hsos}
\begin{array}{lll}
H &=& 
  \frac{1}{2}\sum\limits_i 
    \left\{  
      \sum\limits_{j_1} 
         {\left[ 
           J_1 \delta 
             \left( 
                 \left| h_i  - h_{j_1 } \right| - 1 
             \right)
             + K_1 \delta 
             \left( 
                 \left| h_i  - h_{j_1 }\right|-2\right) 
             \right] 
         }  \right.\\[3ex]
   & & 
     + \sum\limits_{j_2}
           {\left[
              2 J_2 \delta 
                \left(
                  \left| h_i  - h_{j_2 }\right|
                \right) + \left( 2 J_2  + K_2\right)  \delta 
                \left( 
                  \left| h_i  - h_{j_2 } \right| - 3 
                \right) 
           \right] }  \\
   & &
     + 
     \left. 
       J_2 \sum\limits_{j_3} 
         {\left[ 
               \delta 
               \left( \left| h_i  - h_{j_3 } \right| - 2 \right) 
               + \delta 
               \left( 
                  \left| h_i  - h_{j_3}\right| - 4 
               \right) 
         \right]} 
     \right\}+ N J_0, 
\end{array}
\end{equation}

where

\begin{eqnarray*}
J_0 &=&  -\frac{1}{2} \epsilon^{BB}_{1}+ \epsilon_{1}, \\
J_1 &=& -\frac{1}{2} ( \epsilon^{BB}_{2} + \epsilon^{BB}_{3}+ 
2\epsilon^{BB}_{4} )+ \epsilon_{2} + \epsilon_{3}+ \epsilon_{4}, \\
K_1 &=& -\frac{1}{2} ( \epsilon^{BB}_{1} + 2\epsilon^{BB}_{3}+ 
\epsilon^{BB}_{4} )+ \epsilon_{1} + \epsilon_{3}+ \epsilon_{4}, \\
J_2 &=& -\frac{1}{2} \epsilon^{BB}_{4}+ \epsilon_{4}, \\
K_2 &=& -\frac{1}{2} \epsilon^{BB}_{3} + \epsilon_{3}- 
\epsilon_{4}, \\
\epsilon_{i}&=& \epsilon^{AB}_{i}- \epsilon^{AA}_{i},
\end{eqnarray*}
and sums over $j_1$, $j_2$, and $j_3$ denotes summing over first, 
second, and third neighbors of a column at site $i$. In what 
follows, we will treat $J_1$,  $J_2$,  $K_1$, and $K_2$  as model 
parameters. It is worth noting that this   BCCSOS Hamiltonian of  
overlayer-induced faceting at constant coverage can also be used 
to study  a clean bcc(111) surface by setting  
$\epsilon^{AB}_{i}=\epsilon^{AA}_{i}=\epsilon^{BB}_{i}$.  

\subsection{ Energies of  (111), (211), and (110) faces at T=0}\label{s2_stab}

First, we  will check the stability of  ideal surfaces (111), 
(211), and (110) covered with 3, 2, 1 geometric monolayers of A, 
similarly  as it was performed in first principles calculations 
\cite{leung97}. This allows us to estimate values of model 
parameters for reconstruction,  from (111) surface to $\{112\}$  
or to $\{011\}$ faceted surface, under assumption that edge 
energies are neglected.

Using the Hamiltonian Eq.\ (\ref{hsos}) we get the following  
expressions for surface energy per site for ideal face of 
orientation  $(hkl)$

$ \displaystyle E_{111}= E_{r} + 7 J_2$,

$ \displaystyle E_{110}= E_{111}+ 2 J_2 + 2 K_2$,

$ \displaystyle E_{211}= E_{111}+  K_2$, 

where

 $ \displaystyle E_{r}= J_0 + K_1 + 2 J_1$.
 
Thus, the  stability conditions are:

 $K_2 \geq 0$,
 $ J_2 +K_2 \geq 0 \;\;$ for (111) surface,    
 
 $ J_2 +K_2 \leq 0$,
 $ 2 J_2 +K_2 \leq 0\;\;$ for (110) surface,

 $K_2 \leq 0$,
 $ 2 J_2 +K_2 \geq 0\;\;$  for (211) surface.

It easy to see that transformation  from  (111) surface to 
\{211\} facets will be possible for negative values of $K_2$ and 
appropriate positive    $J_2$. To satisfy these conditions,  the 
range  of interactions $\epsilon^{\alpha \beta}_{i}$ should be 
not smaller than $r_4=a \sqrt{11}/2$. On the other hand, it is 
easy to ensure stability of bcc(111) surface at zero coverage ( 
$\epsilon_{i} =0$) by choosing  $\epsilon^{BB}_{3}<0$ and  
$\epsilon^{BB}_{4}<0$. Thus, there is general possibility to 
chose a set of interactions  $\{\epsilon^{BB}_{3},  
 \epsilon^{BB}_{4}, \epsilon_{3},  \epsilon_{4}\}$ in such a way 
that at zero coverage the bcc(111) surface  is stable whereas for 
coverage equal to one physical monolayer the bcc(211) surface 
become stable. 

It is worth noting  that surface energies per site $E_{hkl}$ are 
related to surface energies per surface atom $\sigma_{hkl}$(used 
in first principles calculation \cite{mad99b,leung97}) in the 
following way $ \sigma_{111}= 3 E_{111}$, $ \sigma_{211}=2 
E_{211}$, $\sigma_{110}=E_{110}$, as the number of surface atoms 
is  $N/3$, $N/2$, and $N$, respectively. Moreover, above  
stability conditions correspond to faceting conditions 
\cite{mad99b,leung97} expressed in terms of scaled 
$\sigma_{hkl}$, e.g.,  the conditions $E_{211} \leq E_{111}$,   
$E_{211} \leq E_{110}$ are equivalent to   $3\sigma_{211}/2\leq 
\sigma_{111}$,  $3\sigma_{211}/2\leq 3\sigma_{110}$ for 
transformation $(111)\rightarrow \{112\}$. 

Independence of stability condition on parameters $J_0$, $J_1$, 
and $K_1$ comes from  assumptions on constant number of adsorbate 
atoms A and the restriction on column heights. Therefore we 
neglect these parameters in further analysis what corresponds to  
shifting of the surface energy by $-E_r$. Moreover, by choosing  
$J_2$ as the unit of energy we will work with dimensionless  
quantities:  energy $\tilde H = H/J_2$, temperature  $\tilde T = 
k_BT/J_2$, and parameter $\tilde K = K_2/J_2$ (in what follows, 
the tilde will be omitted).

\section{Simulation of faceting}\label{s3}

We will discus here results of the annealing process investigated 
by Monte Carlo (MC) simulation in canonical ensemble. Let us 
start with some details concerning the simulation method. 
Configurations of BCCSOS model are generated via the classic 
Metropolis algorithm. A new configuration in Markov chain is 
generated from the previous configuration by moving one B atom 
from a site \emph{i} to a site \emph{j}, what is equivalent to 
the following   change of heights: $(h_i, h_j)\rightarrow (h_i-3, 
h_j+3)$. It is important to note that such change of heights is 
possible only if $h_i$ is the local maximum ( with respect to the 
nearest neighbors) and $h_j$ is the local minimum because of the 
restriction on column heights. Let us remember one of the model 
assumption that adsorbate atoms always stay  on the column tops 
as we  study reconstruction  of B surface covered by the 
physical  monolayer of A material. 
\begin{figure}
\centering
\includegraphics[width=7cm]{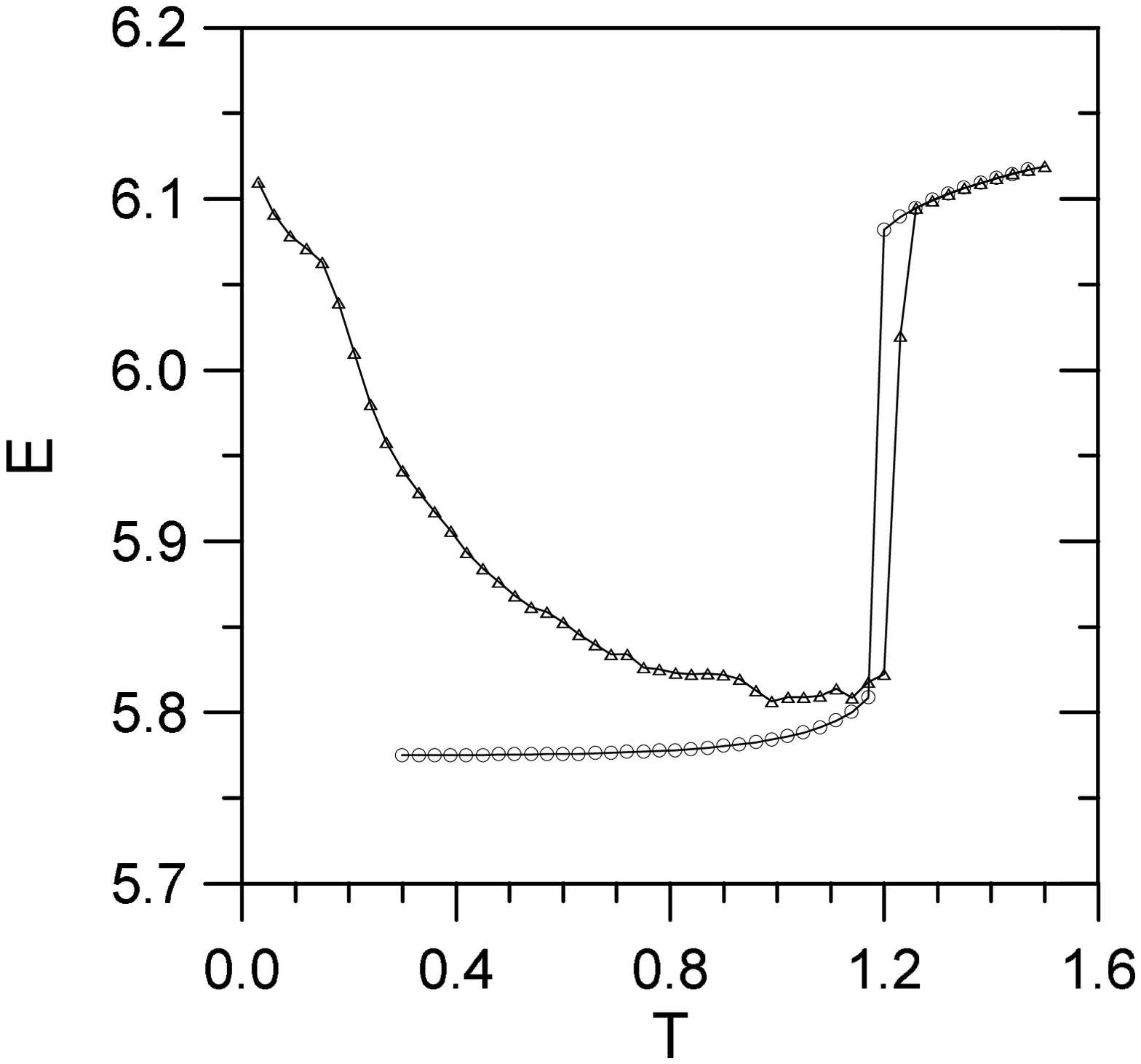}
\caption{ Temperature dependence of the surface energy during the 
warming up cycles (triangles) and the cooling cycles (circles). } 
\label{figure2}
\end{figure} 
 MC simulation were carried out on 
rectangular lattice of  linear size $L_x, L_y$ along x, y 
direction, respectively with periodic boundary conditions (PBC). 
To check the role of PBC additional simulations were performed 
with boundary atoms fixed at positions of  the flat bcc(111) 
surface. The results are consistent with those obtained for 
calculation with PBC. 

First, we investigate properties of the system with coupling 
constant $K=-1.25$ during warming up process. At each 
temperature  $T_i=T_{i-1} + \Delta T$ system spends the same 
annealing time $\tau$ measured in Monte Carlo steps per site. The 
ideal bcc(111) surface is used as a starting configuration at 
$T_0$. We also investigate  properties of the system during 
cooling down the sample. Results of simulations for the following 
parameters: $Lx=192$, $Ly=336$, $\Delta T=0.03$ and $\tau= 
3\times 10^5$ are presented in Fig.\ \ref{figure2}-Fig.\ 
\ref{figure5}. Let us notice that these sizes of the lattice 
correspond to the area of $426\mbox{\AA} \times 431\mbox{\AA}$ on 
the bcc(111) surface with the lattice constant  
$a=3.14\mbox{\AA}$.

\begin{figure}
\centering
\includegraphics[width=8cm]{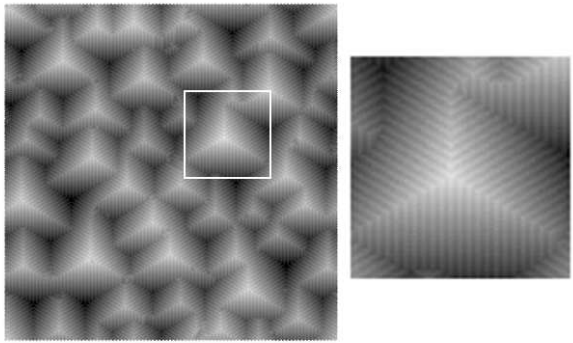}
\caption{Snapshot of the surface at T=0.72 on warming.} 
\label{figure3}
\end{figure}

\begin{figure}
\centering
\includegraphics[width=8cm]{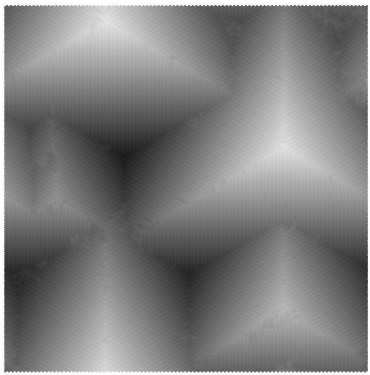}
\caption{Snapshot of the surface at T=1.17 on cooling. } 
\label{figure4} 
\end{figure}

\begin{figure}
\centering
\includegraphics[width=6cm]{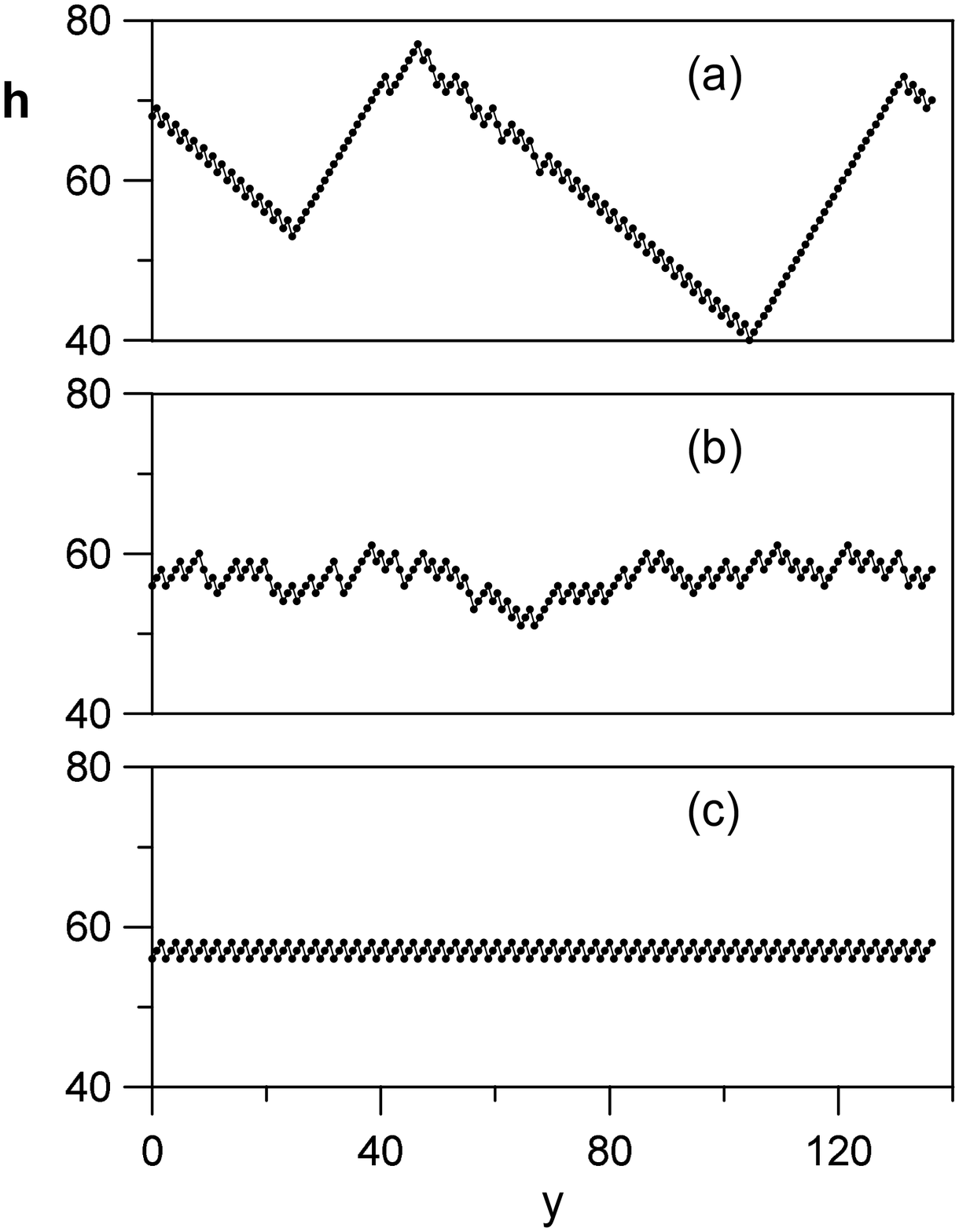}
\caption{ Plot of column height along the y direction for the 
surface: (a) below the phase transition, T=1.17, (b) above the 
phase transition, T=1.50, (c) ideal bc(111) - initial 
configuration. } \label{figure5} 
\end{figure}

The surface formation energy decreases as temperature is elevated 
(see Fig.\ \ref{figure2}) what indicates  reconstruction of the 
surface. At $T\approx 0.4$ very small 3-sided \{211\} pyramids 
are formed on the surface and further warming up causes growth of 
pyramids sizes (see Fig.\ \ref{figure3}). By the \{211\} pyramid 
we mean the pyramid built of (211), (121), and (112) facets.
A big jump of the surface energy is observed at $T= 1.26$ where 
the phase transition to disordered phase occurs. On the other 
hand, discontinuous change of the surface energy is observed  at 
$T= 1.20$ when the system is cooling down. The hysteresis loop 
seen in Fig.\ \ref{figure2} indicates that the phase transition 
is of first order. The surface is undercooled or overheated 
because system rest in metastable state separated from the stable 
state by the free energy barrier. The hysteresis effects, 
dependence of temperature of the phase transition on cooling and 
warming, has been observed in LEED experiment  \cite{song95} for 
Pd/Mo(111).  When system is cooling down we observe formation of 
large pyramids with defected faces and edges just below the 
temperature of the phase transition (see Fig.\ \ref{figure4}). 
Facets of similar shapes and sizes were observed at this 
temperature on warming. During further cooling defects disappear 
and pyramids take nearly ideal shapes at $T\approx 1$. Therefore 
the dependence of surface energy on cooling is very week in the 
ordered faceted phase. 

It is very interesting that in the disordered phase the surface 
is not planar  as suggested in experimental works \cite{song95}. 
The structure of disordered phase contains randomly distributed   
small  facets mainly of $\{112\}$ orientations what is shown in 
Fig.\ \ref{figure5} where the cross section of the disordered 
phase is compared to cross sections of faceted \{211\} and ideal 
bcc(111) surfaces.   Moreover, we calculated the average number 
of atoms  on a facet of orientation (211) which in the disordered 
phase  is  nearly 10. So we will call this phase as disordered 
faceted phase (DFP).  In very narrow temperature range, we 
observed a coexistence of a faceted $\{112\}$ and DFP, e.g., in 
simulation at $T=1.23$ denoted by the triangle  in Fig.\ 
\ref{figure2}. A surface structure in this case looks like 
surface of ordered phase where large pyramids are removed and on 
their place DFP  is present. On the other hand, there are 3-sided 
pyramidal holes built of large facets of $\{112\}$ orientation. 
Sometimes, DFP is located also at the bottom of these holes. It 
is worth noting that coexistence of both faceted and disordered 
phases has been observed \cite{song95}  in LEED experiment in 
Pd/Mo(111) system. However, it was suggested that coexistence is  
due to the inhomogeneity of the surface. 

\begin{figure}
\centering
\includegraphics[width=7cm]{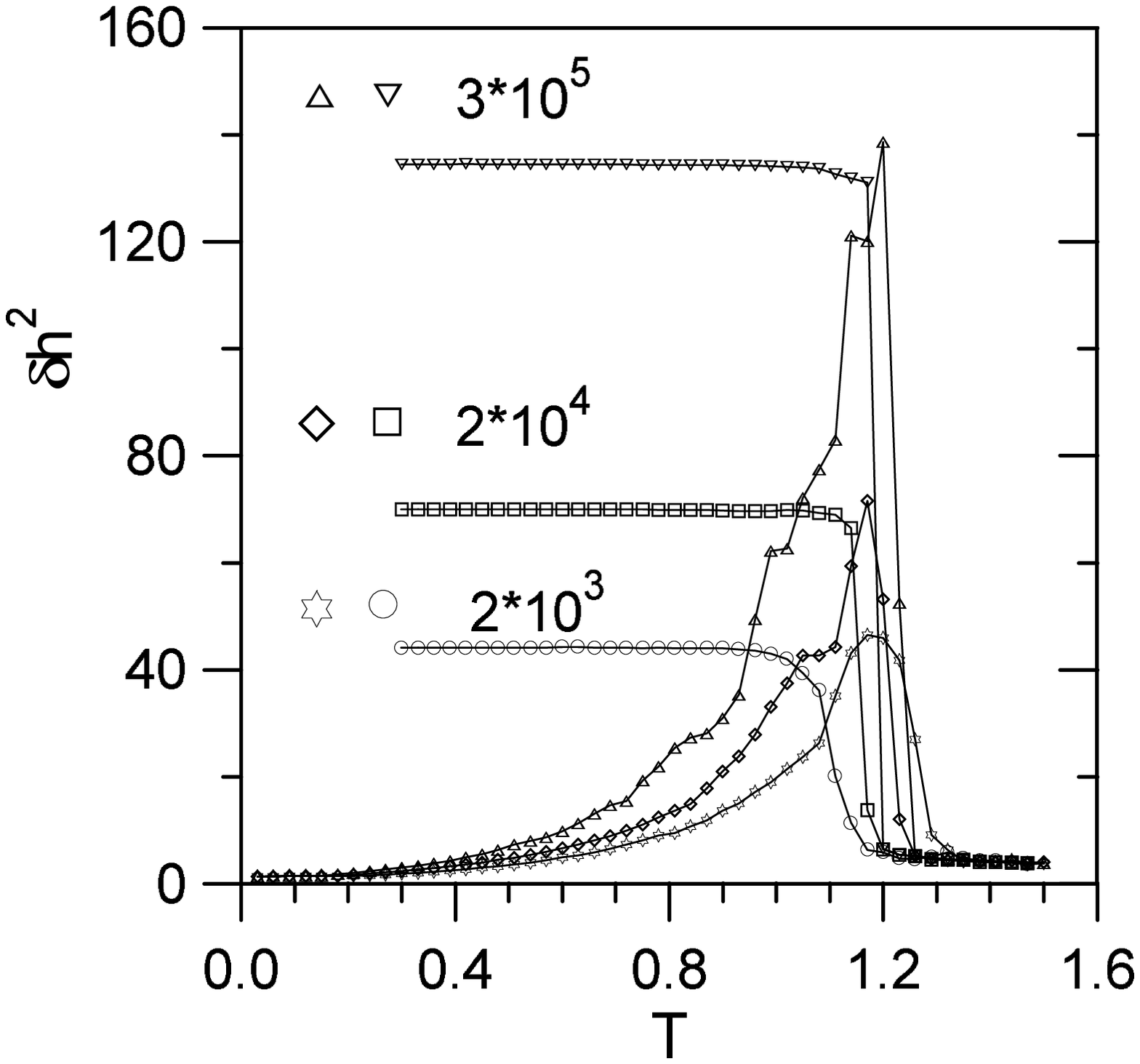}
\caption{ Temperature dependence of the $\delta h^2$ during the 
heating and the cooling cycles for 3 different annealing times.  
Each label contains  symbols  for heating, cooling and the number 
of MC steps} \label{figure6} 
\end{figure}

One of  quantities  measured in MC simulation  is the square mean 
width of the surface,   
\begin{equation}
{\delta h}^2 = 
  \left<   
    \frac{1}{N}    
      \sum\limits_{j} 
         {           
             \left( h_j  - \left<{h}\right>\right)^2                                                       
         }  
    \right>.
\end{equation}
Behaviour of the $\delta h^2$ during warming up and cooling down 
processes allows us to study dependence of mean vertical sizes  
of pyramids on temperature. In Fig.\ \ref{figure6} we present 
results of simulation for 3 different annealing times $\tau= 
2\times 10^3,\;2\times 10^4,\;3\times 10^5$. In all cases the 
hysteresis loop is present near the phase transition and for 
shorter time a larger hysteresis is observed. It is easy to 
explain because it is more probable to overcome the free energy 
barrier in longer time. In the warming up process the size of 
pyramids does not depends on the annealing time $\tau$ up to 
$T\approx 0.5$. When temperature approaches $T\approx 1$ then a 
rapid growth is observed and for longer annealing time  $\tau$  
greater pyramids are formed on the surface. Pyramids reach 
maximal sizes just below the temperature of the phase transition 
where dependence of sizes on annealing time is the strongest. It 
worth noting that dependence of facets size on temperature 
qualitatively agrees with the experimental results \cite{mad94}.
The value of $\delta h^2$ rapidly decreases above the phase 
transition where very weak dependence on temperature is observed. 
In the cooling down process large pyramids are formed just below 
the phase transition temperature, and further decreasing of 
temperature practically does not change the size of pyramids.

We check dependence of surface growth on the initial temperature 
$T_0$ and we find that the behaviour of $\delta h^2$  near the 
transition point is very similar to that in Fig.\ \ref{figure6} 
for 
 process which starts at $T_0=0.3$ or $T_0=1.0$. Finally we 
analyze the influence of the temperature increment,  $\Delta T$, 
on the surface energy and the $\delta h^2$ near the phase 
transition for $\tau=3\times 10^5$. We found that for $\Delta 
T=0.01$ the hysteresis  loop  slightly  increases, there are more 
points where the faceted and disordered phases coexist but it 
seems that sizes of pyramids  do not depend on the temperature 
increment. 

We observe that the growth of the facets on warming base on 
formation of a larger pyramid from several smaller pyramids. 
Probability of such rebuilding of the surface is very small at 
low temperature. This is reflected in the acceptance ratio $f$, a 
fraction of accepted configurations in the Markov chain, which is 
very low at low temperatures, $f \approx 0.002$ below $T=0.3$ and 
$f=0.01$ at $T=0.7$. On the other hand, the acceptance ratio is 
of the same order for $T=1.0$, $f=0.1$,  and close to the phase 
transition,  $f=0.2$, but we observe difference in facets sizes 
at these temperatures. Looking at snapshots of the surface after 
annealing at temperature up to $T\approx 1.1$ on warming, we 
found that  pyramids have nearly ideal shapes (see for example 
Fig.\ \ref{figure3}). We think that all above observation can be 
explained in the following way. The system is passing through 
metastable states - local minima of the free energy - as 
temperature is increased. There are barriers between local minima 
hence the growth of facets depends on the annealing time.

\section{Phase diagram}\label{s4}

\begin{figure}
\centering
\includegraphics[width=7cm]{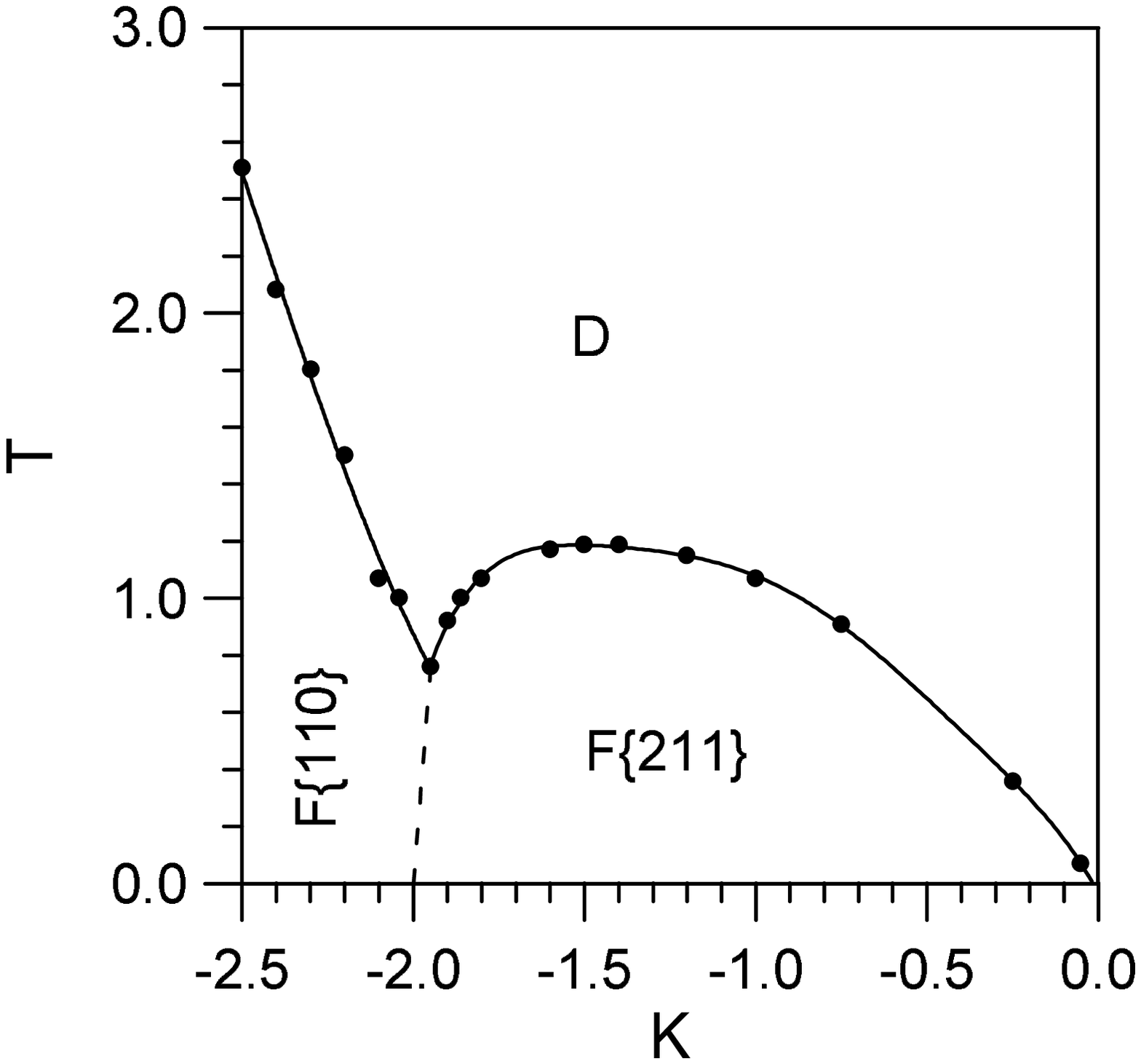}
\caption{ Phase diagram: T versus K. } \label{figure7} 
\end{figure}

The stability analysis of flat (111), (211), and (110) surfaces 
in Sec.\ \ref{s2_stab} shows that at $T=0$ the (211) surface is 
stable for $-2\leq K<0$ whereas for  $K<-2$ the (110) surface is 
stable. Whether these findings are generally correct we can check 
constructing  phase diagram in the (T, K) plane by using MC 
simulations. Most of simulations were performed at constant K 
varying temperature. To check the results some simulations were 
performed at constant temperature. To identify different phases 
we calculated masses of clusters of different facet orientations 
(111), $\{110\}$,  and  $\{112\}$. A cluster of orientation  
(hkl) is  defined in similar way as percolation cluster on the 
(hkl) plane. We also recorded equilibrium configurations at each 
investigated point of the (T, K) plane. We found that for $K<0$ 
the phase diagram ( see Fig.\ \ref{figure7}) contains three phases
  
\begin{enumerate}
  \item Faceted $\{112\}$ phase consisting of 3-sided 
  $\{112\}$ pyramids.
  \item Faceted $\{110\}$ phase consisting of 3-sided 
  $\{110\}$ pyramids.
  \item Disordered faceted  phase.
\end{enumerate}
The surface of the faceted $\{110\}$ phase is built of (110), 
(101), and (011) facets which form 3-sided pyramids on the 
bcc(111) surface. Such surface has been observed experimentally 
in Pd on Ta(111) system \cite{szuk02}.
 The phase transition 
between the disordered phase and one of ordered is of first order 
as surface energy,  the square mean width of the surface $\delta 
h^2$, and the structure factor (see Sec.\ \ref{s5})  change 
discontinuously at the transition point.
The transition temperature between F\{211\} and disordered phase 
is bounded from above by  maximum $T_c=1.20$ at $K=-1.50$. If we 
chose this point to estimate model parameters for Pd/Mo(111) with 
$T_c=850$ K we get $J_2=0.06$ eV, and  $K_2=-0.09$ eV.  On the 
other hand the transition temperature from  F\{110\} to the 
disordered phases is increasing function of $|K|$  and it might 
reach large value. The present model does not include the 
desorption, thus we can expect that  transition temperature might 
be above the desorption temperature in some systems especially 
where the faceted F\{110\}  phase occurs. In such  cases the 
phase transition to the disordered faceted phase could not be 
observed but rather deconstruction induced by desorption would be 
expected.

There is difficult to study phase transition between ordered 
phases using MC simulation because for $K\approx -2$ we observed 
coexistence of two types of facets $\{011\}$  and  $\{112\}$. The 
mixed phase does not contain separated pyramids of two types but 
mixture of facets, e.g.,  on the (211) facet a smaller (011) 
facet can occur. 

\section{Simulation of LEED patterns}\label{s5}

Low  energy electron diffraction (LEED) experiments play an 
important role in investigation of structures and phase 
transitions of faceted surfaces \cite{mad94, song95}. Very 
recently, Song et al.\ \cite{song95} where able to study  LEED 
patterns of Pd/Mo(111) system in high temperatures up to 1200K. 
They demonstrated the existence of reversible planar/faceted 
phase transition with transition temperature different for 
heating cycles ($T\approx 870K$)  and for cooling cycles 
($T\approx 830K$). 

\begin{figure}
\centering
\includegraphics[width=6cm]{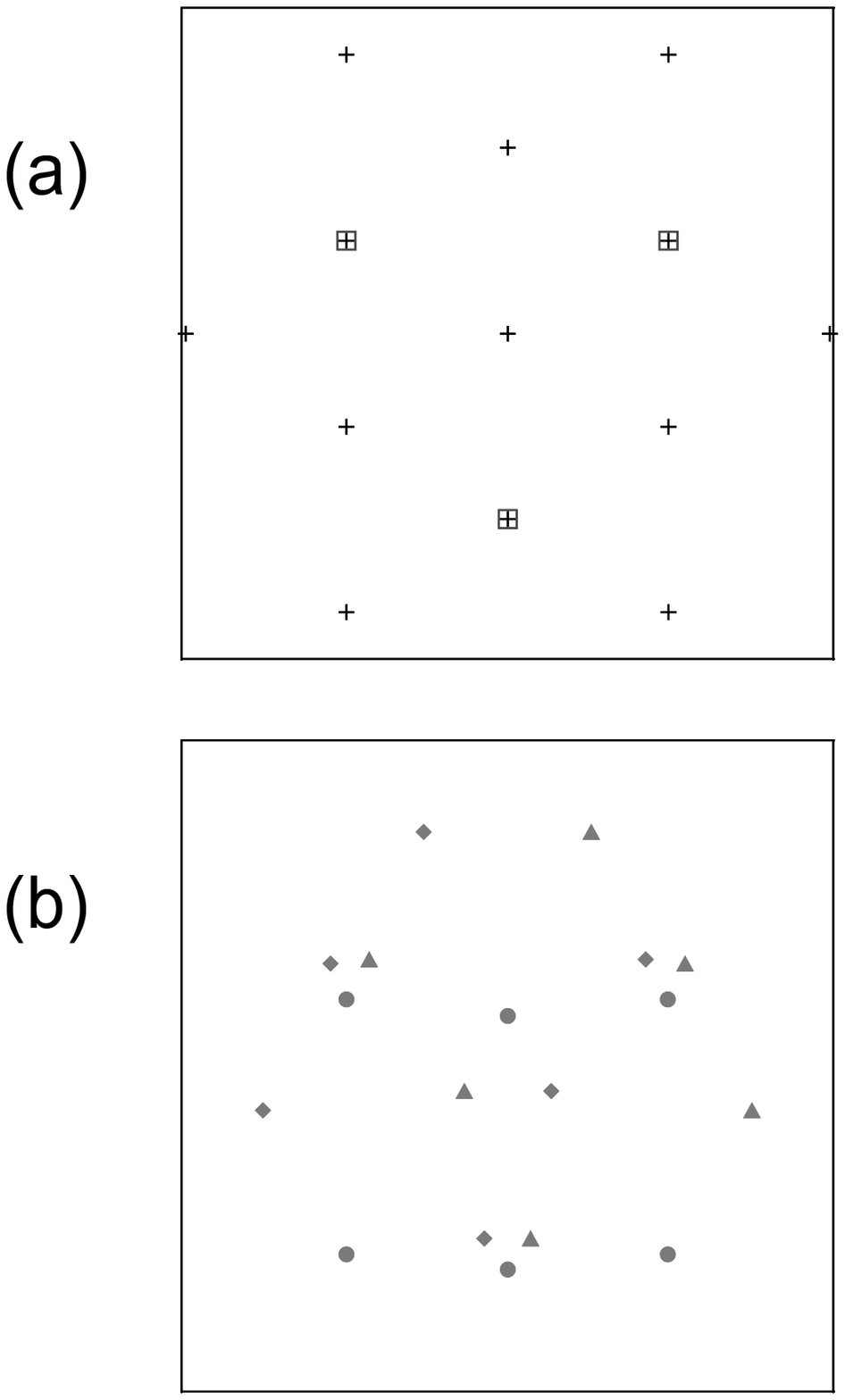}
\caption{ LEED patterns of (a) disordered phase where spots are 
represented by squares and also spots of ideal bcc(111) surface 
(crosses) are shown, (b) faceted \{211\} phase where circles, 
diamonds, and triangles denote spots coming from diffraction on 
facets (121), (211), and (112), respectively. } \label{figure8}
\end{figure}

In this section we employee the BCCSOS model to analyze the 
temperature dependence of LEED patterns. The diffracted intensity 
is proportional, in the kinematic approximation, to the structure 
factor 

\begin{equation}
S({\bf k}) = 
  {\left<  
    \left|        
      \sum\limits_{j} 
         {           
            \exp(i {\bf k  R}_j)                                                       
         }  
     \right|^2
   \right>},  
\end{equation}
where  summing is over surface atoms, ${\bf k}= {\bf k}_f-{\bf 
k}_i$, and  ${\bf k}_i$, ${\bf k}_f$ is the wave vector of 
incident electron, scattered electron, respectively. Position of 
surface atoms  ${\bf R}_j$ is expressed by column height 
$h_{r_j}$ in site  ${r_j}=(x_j, y_j)$   in the following way 
${\bf R}_j= a(3 \sqrt{2} x_j, \sqrt{6} y_j, \sqrt{3} h_{r_j})/6$. 
In a SOS model not all atoms from column tops are surface atoms. 
Therefore in calculation of diffracted intensity sometimes the 
shadowing factor is used \cite{tosat96}. In this paper we 
consider the contribution to $S$ only from surface atoms. As a 
surface atom in any configuration we choose the atom from the top 
of column which has  at least four lower nearest neighbors. In 
the case when number of higher columns is equal to 3 then atom is 
regarded as surface atom if is not surrounded by the higher 
neighbors. Such case occurs for atoms on   \{110\} facets. We 
will calculate  structure factor only for a set of ${\bf k}_f$, 
chosen  in such a way to describe diffraction on faces (111) and 
\{211\}. For each of these faces we construct a pair of inverse  
lattice vectors ${\bf g}_1, {\bf g}_2$  parallel to the face, 
e.g., for (121) ${\bf g}_1=2\pi/a(\sqrt{2}/2, 0, 0)$ and  ${\bf 
g}_2=2\pi/a(0, 4\sqrt{6}/9, 2\sqrt{3}/9)$. Then, the wave vectors 
of scattered electrons, ${\bf k_f}$,   for given face are chosen  
to satisfy condition of constructive interference  $({\bf 
k}_f-{\bf k}_i)_{\parallel} =l{\bf g}_1+m{\bf g}_2 $, where $l, 
m=0, \pm 1, \ldots$ and the  symbol $\parallel$  denotes vector 
component parallel to the face. 

We calculated structure factors  for the wave vector of the 
incident electron, $ k_i =2\pi/a( 0, 0, 1.43)$, normal to the 
bcc(111) surface. Thus energy of incident electrons is equal to 
31 eV  for $a=3.14\mbox{\AA}$,  similarly as in LEED experiment 
for Pd/Mo(111) \cite{song95}. Figure~\ref{figure8} shows   LEED 
patterns obtained in computer simulation for the model parameter 
$K=-1.25$. In the ordered  phase  \{211\}  we see besides 3 
clovers (groups of 3 spots) observed experimentally, the forth 
clover in the center of the image, and six single spots more 
distant from the center. The pattern of disordered phase  
consists of three spots which positions correspond to the centers 
of external clovers in the diffracted pattern of the faceted 
phase. Apparent lack of spots  from bcc(111) surface confirms 
that disordered phase has not the planar structure. 

\begin{figure}
\centering
\includegraphics[width=7cm]{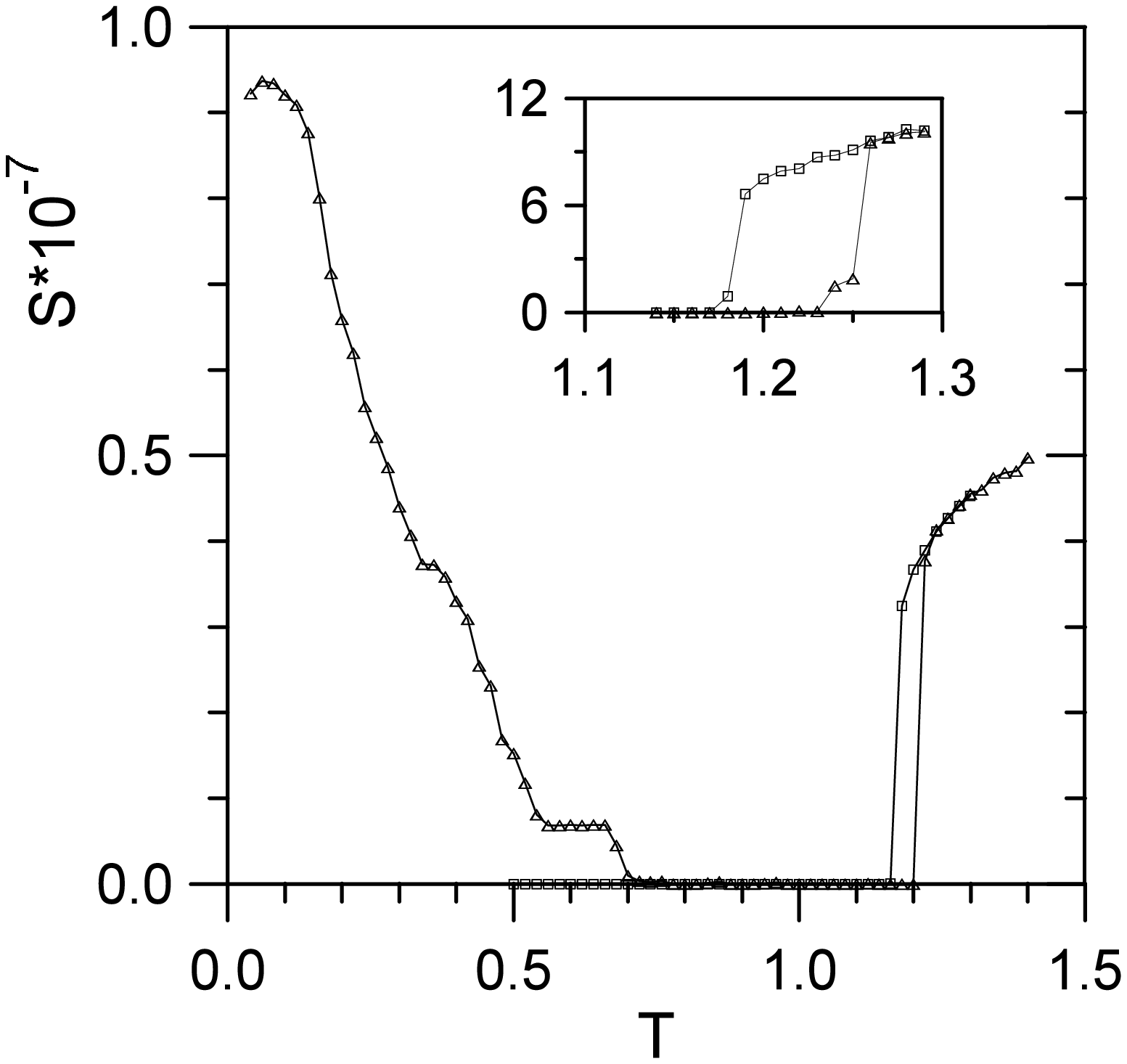}
\caption{ Temperature dependence of the structure factor  for the 
wave vector corresponding to a spot of the disordered phase on 
warming (triangles) and on cooling  (squares). The hysteresis 
loop for large lattice $Lx=192$, $Ly=336$, is shown in the inset. 
} \label{figure9}
\end{figure} 
Temperature dependence of the structure factors were calculated 
using MC simulation and changing temperature in the way described 
in Sec.\ \ref{s3}. Most of simulations were performed on the 
lattice with   $Lx=90$, $Ly=132$. The spots of disordered phase 
have the same diffracted intensities. Their temperature 
dependence (see Fig.\ \ref{figure9})  on warming and on cooling 
shows the presence of hysteresis loop in behaviour of $S(T)$ 
close to the phase transition what is in agreement with 
experimental results  \cite{song95}. The hysteresis loop 
calculated for large lattice with $Lx=192$, $Ly=336$ is shown in 
the inset in Fig.\ \ref{figure9}. There are several points where 
the structure factor takes intermediate values because at these 
temperatures   coexistence of faceted and disordered phases has 
been observed. More coexistence points where found  on warming 
than on cooling. We observe also discontinuity of the structure 
factor at the transition temperature. Thus hysteresis and 
discontinuity of the structure factors are another argument for 
the first-order phase transition between faceted and disordered 
phases. During the process of warming up  the system the S 
decreases with temperature and becomes very small when on the 
surface large enough \{211\} facets   are formed. 

\begin{figure}
\centering
\includegraphics[width=7cm]{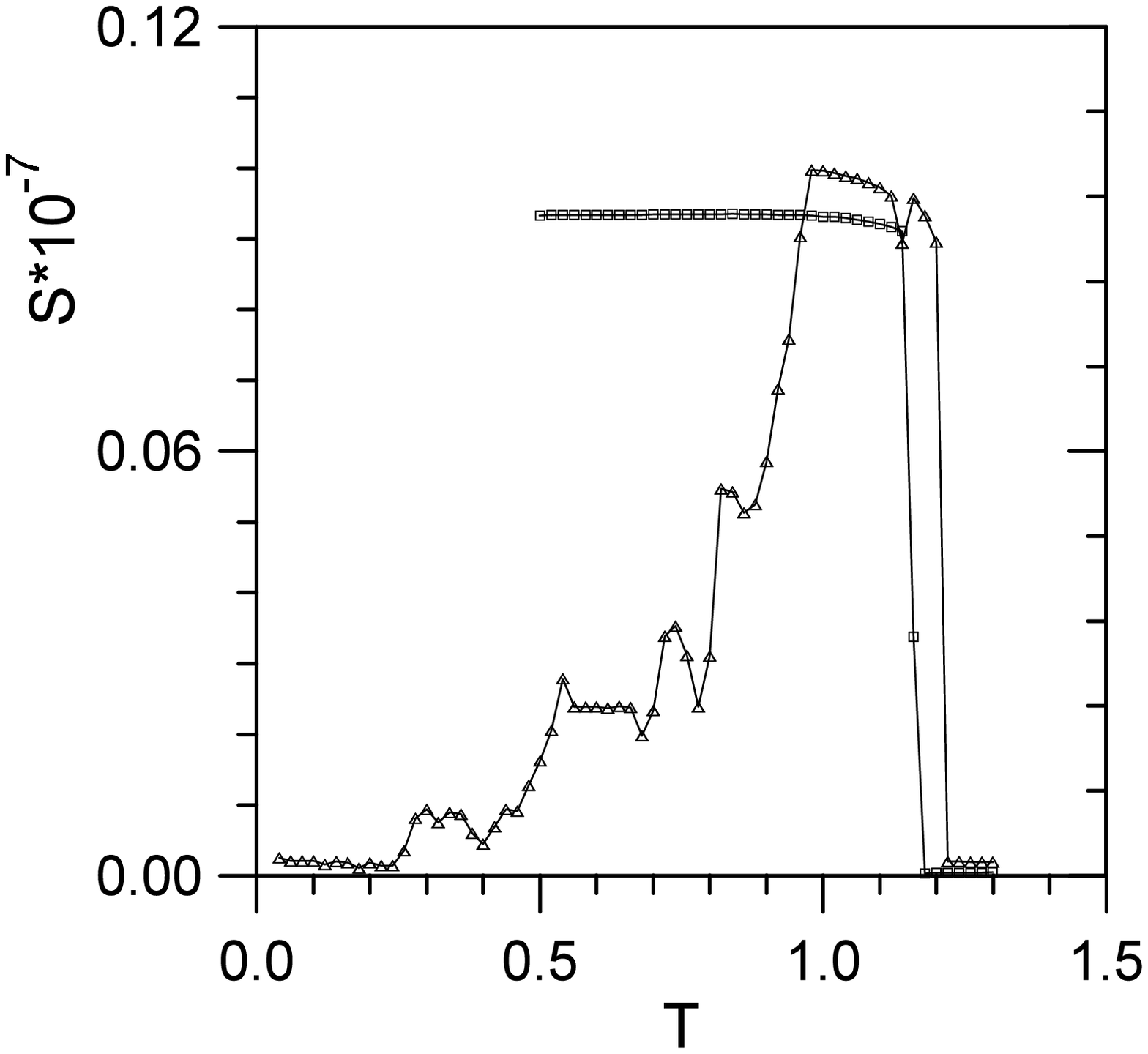}
\caption{ Temperature dependence of the  structure factor for the 
wave vector corresponding to a leaf of  the right clover  on 
warming (triangles) and on cooling (squares). } \label{figure10}
\end{figure} 
The diffracted intensities of a spot in the ordered phase was 
calculated for ${\bf k}_f$ fulfilling the condition of 
constructive interference and also for vectors from a very small  
surroundings of ${\bf k}_f$. Then maximal value of S were chosen 
to represent the intensity of spot at this ${\bf k}_f$. We use 
such procedure because the surface does not consist of ideal flat 
\{211\} facets during warming up and cooling processes, therefore 
spots can change shapes, intensities, and positions. The 
diffracted intensities of spots  of faceted phase are different 
because areas of facets of different orientation are not equal. 
Moreover, their dependence on temperature during warming the 
system shows irregular behaviour (see Fig.\ \ref{figure10})  
because of facets grow i.e., from several small facets a bigger 
one arises.   We observe rapid growth of S at temperature 
$T\approx 0.9$    when  large pyramids are formed on the 
surface.  During the cooling cycles, the S reaches nearly maximal 
value just below transition temperature what confirms formation 
of large \{211\} facets. The maximal values of S in the faceted 
phase are nearly ten times smaller than  in the disordered 
phase   because in  former case only $\sim 1/3$ surface atoms  
contribute to S, i.e., surface atoms placed on facets of the same 
orientation.

\section{Discussion}
We have presented here the BCCSOS model to study overlayer 
induced faceting. Although it might seem  that model is very 
simple it gives many results in agreement with experiments. It 
has been shown that bcc(111) surface covered by a physical 
monolayer undergoes reconstruction upon annealing. We obtained 
two types of reconstructed surfaces: faceted \{211\} covered by 
3-sided pyramids as observed in Pd/Mo(111) \cite{mad96} and 
faceted \{011\} surface found in Pd/Ta(111) \cite{szuk02}. The 
sizes  of facets depend on annealing  temperature  and annealing 
time.  At high temperature the phase transition  to disordered 
phase occurs. We have shown that LEED diffraction patterns can be 
calculated, in kinematic approximation, in warming up  and 
cooling down processes for different  electron energies. In 
dependence of diffracted intensities on temperatures  the 
hysteresis effect was observed close to  temperature of the phase 
transition.

On the other hand, we can investigate  some properties not 
reported in experimental  papers. First of all the phase 
transition can be studied in details using this model. It is 
shown that the phase transition is of first order because 
quantities such as surface energy, the structure factor change 
discontinuously as temperature reaches the transition point. 
Moreover the hysteresis effects  are observed when system is 
warming up and then cooling down. We can analyze the structure of 
the surface in disordered phase as well as in  the coexistence of 
the faceted and disorder phase close to the transition 
temperature. We found that the disordered phase consists  of many 
small \{211\}  facets randomly distributed. Hence it is not 
planar  bcc(111) surface.  It would be interesting to check 
experimentally  this prediction.  It was suggested \cite{song95} 
that coexistence of faceted and disordered phases is  due to the 
inhomogeneity of the adsorbate coverage but we have demonstrated 
that the coexistence could also exist for homogeneous coverage. 

It seems that this model can be applied to study  reconstruction 
of curved bcc surfaces where reconstructed surface has different 
structure than faceted bcc(111) surface. In case of Pd deposited 
on a needle-shaped tungsten \cite{sczep02},    step like \{211\} 
microfacets  has been observed.  Using the present  model we have 
not obtained the faceted  phase with large \{211\} pyramids and 
small \{011\} pyramids as observed in Pd/W(111) \cite{mad99b} for 
prolonged annealing time. This might be due to lack of density 
fluctuation. So we are going  to extend the  model including  
dependence on coverage. 
 The most general extension should base on replacing of the interactions 
$\epsilon^{\alpha \beta}_{i}$ by many-body potentials. To 
construct such potentials, e.g., for Mo-Pd, W-Pd systems, the 
results of the first-principle calculations could be used 
\cite{zhang98}.

\subsection*{Acknowledgements} I would like to thank prof.\ Jan 
Ko{\l}aczkiewicz and dr Andrzej Szczepkowicz for  discussions.  
This work was supported by the Polish State Committee  of 
Scientific Research (KBN) Grant No 2 P0 3B 107 19.

\end{document}